\documentclass[aps,prb,twocolumn,superscriptaddress,floatfix]{revtex4}
\usepackage{epsfig}
\begin{document}

%%% Slightly higher pages to make it fit on 4 pages
\advance\paperheight by 3 true mm
\advance\textheight by 3 true mm

\title{Thermal conductivity of anisotropic and frustrated spin-$1/2$ chains}

\author{F. Heidrich-Meisner}\author{A. Honecker}
\affiliation{Technische Universit\"at Braunschweig, Institut f\"ur Theoretische
  Physik, Mendelssohnstr.\ 3, 38106 Braunschweig, Germany}
\author{D. C. Cabra}
\affiliation{Departamento de F\'{\i}sica, Universidad Nacional de La Plata,
  C.C.\ 67, (1900) La Plata, Argentina}
\affiliation{Facultad de Ingenier\'{\i}a, Universidad de Lomas de Zamora,
  Camino de Cintura y Juan XXIII, (1832) Lomas de Zamora, Argentina}
\author{W. Brenig}
\affiliation{Technische Universit\"at Braunschweig, Institut f\"ur Theoretische
  Physik, Mendelssohnstr.\ 3, 38106 Braunschweig, Germany}

%\email[F. Heidrich-Meisner]{f.heidrich-meisner@tu-bs.de}

\date{August 14, 2002; revised: September 18, 2002}
%--------------------------------------------------------------------------
\begin{abstract}
We analyze the thermal conductivity of anisotropic and frustrated
spin-$1/2$ chains using analytical and numerical techniques. This includes
mean-field theory based on the Jordan-Wigner transformation, bosonization,
and exact diagonalization of systems with $N\leq 18$ sites. We present results
for the temperature dependence of the zero-frequency weight of the conductivity
for several values of the anisotropy $\Delta$. In the gapless regime, we show
that the mean-field theory compares well to known results and that the
low-temperature limit is correctly described by bosonization. In the
antiferromagnetic and ferromagnetic gapped regime, we analyze the temperature
dependence of the thermal conductivity numerically. The convergence of the
finite-size data is remarkably good in the ferromagnetic case. Finally, we
apply our numerical method  and mean-field theory to the frustrated chain where we find
a good agreement of these two approaches on finite systems.
Our numerical data do not yield evidence for a diverging thermal conductivity
in the thermodynamic limit
in case of the antiferromagnetic gapped regime of the frustrated chain.
\end{abstract}
%--------------------------------------------------------------------------

\maketitle

\begin{figure}[t]
\centerline{\epsfig{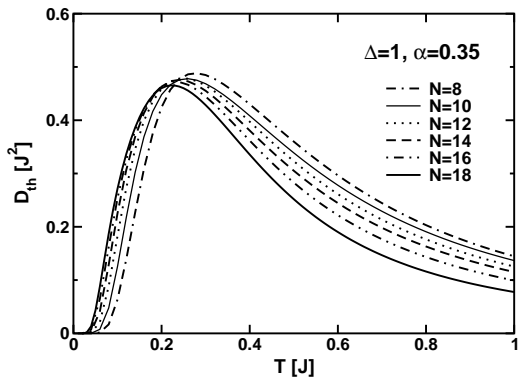}}
\caption{Thermal Drude weight $D_{\mathrm{th}}(T)$ for the frustrated chain with $\Delta=1,\alpha=0.35$ for $N=8,\dots,18$
sites.\label{fig:3}}
\end{figure}

%--------------------------------------------------------------------------
{\bf Introduction - }
Transport properties of low-dimensional spin systems have attracted recently interest
both from the experimental and theoretical side. A particular motivation comes from
the observation that magnetic excitations of one-dimensional spin systems significantly
contribute to the thermal conductivity which is manifest in many experiments on materials such as the
spin-ladder system\cite{solo00,hess01,kudo01}  $\mbox{(Sr,La,Ca)}_{14}\mbox{Cu}_{24}\mbox{O}_{41}$ and
the spin-chain compounds $\mbox{SrCu}\mbox{O}_2$ and $\mbox{Sr}_2\mbox{Cu}\mbox{O}_3$\cite{solo01}.
Assuming elementary excitations to carry the thermal current and using a relaxation time ansatz
for their kinetic equation
one finds extremely large mean-free paths being, for example, of
the order of 1000\AA\enspace in  $\mbox{La}_5\mbox{Ca}_{9}\mbox{Cu}_{24}\mbox{O}_{41}$\cite{hess01}.
Although the magnitude   of the mean-free path is currently an issue of intense discussion,
the question arises whether heat transport in low-dimensional spin systems is
ballistic, i.e., whether intrinsic scattering of magnetic excitations is ineffective
to render the thermal conductivity finite.
From the theoretical point of view this issue is related to the value of the so-called (thermal)
Drude weight\cite{zotos97} $D_{\mathrm{th}}$ which is the zero-frequency weight
 of the thermal conductivity $\kappa$.
A nonzero value of $D_{\mathrm{th}}$ corresponds to a diverging thermal conductivity.
This scenario is trivially realized if the energy-current operator is a conserved quantity,
which is the case for the spin-$1/2$ Heisenberg chain\cite{niemeyer71,zotos97}.
For a number of other models like  the  frustrated
chain, the dimerized chains or the spin ladder the energy-current operator is not conserved
and the question of nonzero $D_{\mathrm{th}}$ is a challenging topic. \\
\indent In this paper, we  establish various numerical and analytical techniques
to analyze the thermal Drude weight and to compute the temperature dependence of $D_{\mathrm{th}}(T)$.
We study the model Hamiltonian $H=\sum_{l} h_l$ with the local energy-density given by
\begin{equation}
h_l= J   \lbrace (S_l^+S_{l+1}^-+\mbox{H.c.})/2+{\Delta }S_l^{z}S_{l+1}^{z}+\alpha \vec{S}_l\cdot\vec{S}_{l+2}\rbrace
\label{eq:1}.
\end{equation}
The $XXZ$ model ($\alpha=0$) is integrable whereas for nonzero frustration the
model becomes nonintegrable. Recently, Kl\"umper and Sakai \cite{kluemper02}
obtained $D_{\mathrm{th}}(T)$ for $\alpha = 0$ and $0\leq\Delta\leq 1$
by using the Bethe ansatz\cite{kluemper} which allows us to test our approaches
in this regime.
Exact diagonalization of finite systems up to $N=14$ sites
has been applied by Alvarez and Gros \cite{gros02} to investigate
$D_{\mathrm{th}}(T)$ for the isotropic Heisenberg chain (i.e.,\ $\Delta=1$),
the frustrated chain, and the spin ladder. Our numerical analysis
goes beyond this by allowing $\Delta \ne 1$ and extension to larger
systems with $N\leq 18$ sites. \\
\indent
%----------------------------------------------------------
{\bf Thermal conductivity - }
The thermal conductivity $\kappa$ is defined by
$\langle j\rangle = -\kappa \nabla T$
and is given by the following expression\cite{mahan}:
\begin{equation}
\kappa(\omega) = \beta \int_0^{\infty} dt\, e^{-i\omega t}
         \int_0^{\beta} d\tau \langle j(-t-i\tau)j\rangle .
\label{eq:12}\end{equation}
$j$ is the energy-current operator\cite{current} and $\beta=1/T$ is the
inverse temperature.  The current operator satisfies the equation of continuity:
$\partial_t h_{l} =i[H,h_{l}]=-( j_{l+1}-j_{l})$.
For exchange interaction of arbitrary range, i.e.,
$\lbrack h_{l\pm m},h_l\rbrack\not= 0$   for $m\leq m_0$, this implies
\begin{equation}  j_l=i\sum_{m,n=0}^{m_0-1}\lbrack h_{l-m-1},h_{l+n}\rbrack.
\label{eq:5a}\end{equation}
In our case we have $m_0=2$ (see Eq.\ (\ref{eq:1})) leading to
\begin{equation}
j=\sum_l j_l=i\sum_l \lbrack h_{l-2}+h_{l-1}, h_{l}+h_{l+1}\rbrack.
\label{eq:10}\end{equation}
Note that the current operator derived from Eq.\ (\ref{eq:10})
includes the proper limiting form  for $\alpha\gg 1 $
where one  recovers  the current operators of two decoupled chains.\\
\indent The conductivity $\kappa(\omega)$ may be decomposed according to
$\mathrm{Re}\,\kappa(\omega)=D_{\mathrm{th}}(T) \,\delta(\omega)+\kappa_{\mathrm{reg}}(\omega)$
into a singular part at zero frequency and a regular part $\kappa_{\mathrm{reg}}(\omega)$.
The quantity of interest is the thermal Drude weight $D_{\mathrm{th}}(T)$ which
can be computed via (see, e.g.,\ Ref.\ \onlinecite{zotos97})
\begin{equation}
D_{\mathrm{th}}(T) =
   \frac{\pi \beta^2}{Z\,N} \sum_{m, n\atop E_m=E_n}e^{-\beta E_m}
      |\langle m | j|n\rangle|^2 .\label{eq:13}\end{equation}
$Z$ is the partition function and $N$ the number of lattice sites.
Note that we exclusively use periodic boundary conditions.
If $j$ is a conserved quantity, the conductivity reduces to
$\kappa(\omega)=D_{\mathrm{th}}(T)\delta(\omega)$ and expression (\ref{eq:13})
simplifies to  $D_{\mathrm{th}}=\pi \beta^2 \langle j^2 \rangle/N$. \\
%------------------------------------------------------------------------------------------
\begin{figure}[t]
\centerline{\epsfig{figure=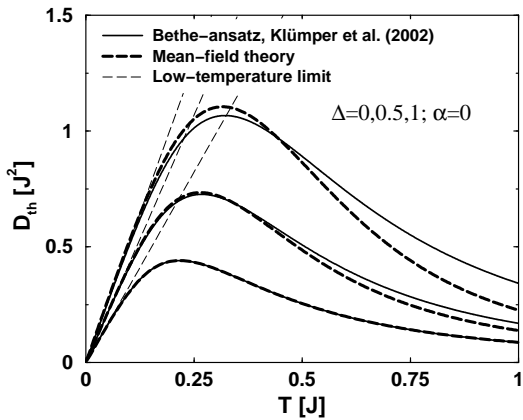,width=0.38\textwidth}}
\caption{
$D_{\mathrm{th}}(T)$ for different values of the anisotropy $\Delta=1,0.5,0$
(top to bottom) and zero frustration. Dashed lines denote results obtained by
Jordan-Wigner transformation and mean-field treatment of the interaction term.
Bethe-ansatz results from Ref.\ \onlinecite{kluemper02} are included in
the figure (thick solid lines). Thin dashed lines show the exact result for
the low-temperature limit (see Eq.\ (\ref{eq:17d})).\label{fig:1}}
\end{figure}
\indent
{\bf Mean-field theory - }
Using a Jordan-Wigner transformation\cite{mahan} the spin operators are mapped to spinless fermionic
operators $c_l^{(\dagger)}$. In the case of $\Delta=0$ and zero frustration $\alpha$ the corresponding
Hamiltonian is diagonal in momentum space and
reads $H=\sum_k \epsilon_kc^{\dagger}_k c_k$ with a tight-binding dispersion $\epsilon_k=-J \cos(k)$. A nonzero
value of $\Delta$ or $\alpha$ leads to a four-fermion interaction term that can be treated approximately by
Hartree-Fock (for details, see, e.g.,\ Ref.\ \onlinecite{brenig97}),
resulting in a renormalization of
$\epsilon_k$ to $\tilde \epsilon_k=-J(1+2A(\Delta-2\alpha))\cos(k)$.
The parameter $A=\frac{1}{\pi}\int_{0}^{\pi}dk\, \cos(k)f(\tilde\epsilon_k)$
has to be determined self-consistently where $f(\epsilon)=1/(\mathrm{exp}(\beta\epsilon)+1)$ is the Fermi function.
Using Eq.\ (\ref{eq:13}) the thermal conductivity $\kappa(\omega)=D_{\mathrm{th}}(T)\delta(\omega)$
can be computed directly.
Here we focus on the case of $0\leq\Delta\leq 1$ and
$\alpha<\alpha_{\mathrm{crit}}$\cite{hase93}, which is known to exhibit gapless spinonlike excitations.
 Results for $D_{\mathrm{th}}(T)$
of the $XXZ$ model are shown in Fig.\ \ref{fig:1} (thick dashed lines).
For comparison Bethe-ansatz results\cite{kluemper02}  are
included in the figure (solid lines).  In the case of $\Delta=0$,
no approximations are necessary in
 the Jordan-Wigner approach   and  consequently, Jordan-Wigner and Bethe-ansatz results are identical.
  For $\Delta>0$, the main observation is that the mean-field theory  produces qualitatively the
right  picture of the temperature dependence of $D_{\mathrm{th}}(T)$.
Both the slope of $D_{\mathrm{th}}\sim T$
at low temperatures and the position of the maximum are  well predicted.
Deviations at high temperatures are due to the neglect of many-particle excitations in the mean-field
approximation.\\
%------------------------------------------------------------------------------------------
\indent
{\bf Bosonization - }
In the continuum limit the physics of the anisotropic spin-$1/2$ chain at low
energies is described by the Luttinger-liquid Hamiltonian\cite{schulz}
\begin{equation}
H = {1 \over 2} \int dx \left( v K (\partial_x \Theta)^2
+ {v \over K} (\partial_x \phi)^2 \right) ,
\label{eq:16}
\end{equation}
where $\phi$ is a bosonic field in $1+1$ dimensions and $\Theta$ is the
dual field $\partial_x \Theta= \frac{1}{K}\partial_{\tau}\phi$.
$K$ is the Luttinger parameter and $v=(J \pi/2)\frac{\sin{\gamma}}{\gamma}$
is the spinon velocity where
the anisotropy is parametrized via $\Delta=\cos(\gamma)$ here.
The local current operator $j(x)$ is again given by the equation of continuity:
$\partial_x\,j(x)=-\partial_t h(x)$.
We obtain
\begin{equation}
j= v^2\int dx\,
\partial_x\phi(x) \partial_{x}\Theta(x).\label{eq:17}
\end{equation}
The Drude weight follows from $D_{\mathrm{th}}=\pi\beta^2\langle j^2\rangle/N$.
Thus we have to evaluate the two-point function
$\langle j(x,\tau)j(0,0)\rangle $, $\tau$ being the time variable.
The computation is  similar to the procedure for the susceptibility
in Ref.\ \onlinecite{eggert94}.
We change to coordinates $z = v\tau+ix$ and $\bar z = v\tau-ix$.
By decomposing $\phi(z,\bar z)=\varphi(z)+\bar \varphi(\bar z)$ into its
chiral parts and using the respective two-point functions such as
$\langle \varphi(z)\varphi(w)\rangle =-({K}/{4\pi})\mathrm{ln}(z-w)$
we obtain
\begin{equation}\langle j(x,\tau)j(0,0)\rangle =-2 \frac{v^2}{(4\pi)^2}\left(
\frac{1}{z^4}+\frac{1}{{\bar z}^4}\right).\label{eq:17c}\end{equation}
Before performing the space integration the imaginary time direction is
compactified by mapping the plane ($z$) into the strip ($\zeta$) using
$z(\zeta)=\mathrm{exp(2\pi\zeta/\beta)}$ leading to the replacement
$v\tau\pm ix\to( v\beta/\pi) \sin(\pi\frac{v\tau\pm ix}{v\beta})$ in
Eq.\ (\ref{eq:17c}). After the change of variables
$u=\tan(\pi \tau/\beta);w=-i\tan(i\pi x/(v\beta))$ we finally find
\begin{equation} D_{\mathrm{th}}(T)=\frac{\pi^2}{3}\,v\,T.
\label{eq:17d}\end{equation}
This coincides with Kl\"umper's and Sakai's analytic
expression\cite{kluemper02} for the low-temperature limit
of the $XXZ$ model if the velocity $v$ is equal to
$v=(J \pi/2)\frac{\sin{\gamma}}{\gamma}$. However, the result
is more generally valid for models with the continuum limit given by the
Luttinger-liquid Hamiltonian.\\
%------------------------------------------------------------------------------------------
\begin{figure}[t]
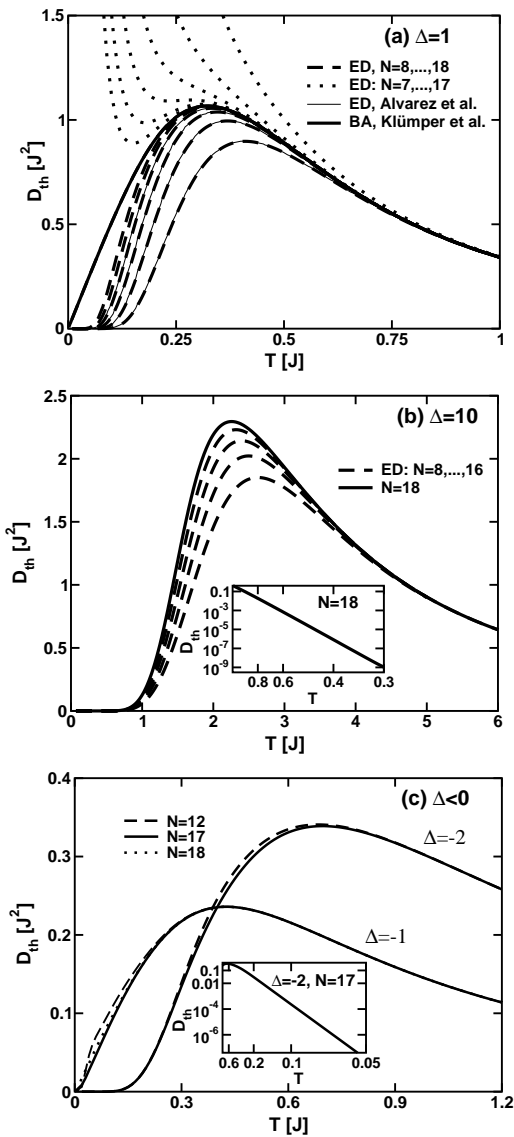

\centerline{\epsfig{figure=figure2.eps,width=0.36\textwidth}}
\vspace{0.18cm}
\centerline{\epsfig{figure=figure3.eps,width=0.37\textwidth}\hspace{0.2cm}}
\vspace{0.18cm}
\centerline{\epsfig{figure=figure4.eps,width=0.375\textwidth}}
\caption{Exact diagonalization (ED) for the $XXZ$ chain: (a) Isotropic chain $(\Delta=1)$.
Dashed(dotted) lines denote even(odd)-numbered systems for
$N\leq 18$ sites.
Bethe-ansatz results by Kl\"umper and Sakai\cite{kluemper02} are
included in the plot (thick solid line). Thin solid lines show ED results by Alvarez and
Gros\cite{gros02} for $N=8,10,12,14$. (b) Antiferromagnetic, gapped regime ($\Delta=10$). ED for
$N=8,\dots,16$ (dashed lines), $N=18$ (solid line).
(c) Ferromagnetic regime ($\Delta=-1,-2$). Note: the thermodynamic limit is  reached for $N\approx 17$.
The insets of (b) and (c) display the exponential suppression of $D_{\mathrm{th}}(T)$ at low temperatures for $\Delta=10$
and $\Delta=-2$ (in the insets, vertical axes are scaled logarithmically, horizontal axes reciprocally).}\label{fig:2}
\end{figure}
\indent {\bf Exact diagonalization (ED) - }
In this part we present our results for $D_{\mathrm{th}}(T)$
obtained by exact diagonalization for finite
systems with $N\leq 18$. We start with the discussion of
different values of the anisotropy $\Delta$ at zero frustration.
Figure \ref{fig:2} shows $D_{\mathrm{{th}}}$ for the isotropic
case $\Delta=1$, for a gapped, antiferromagnetic
system ($\Delta=10$) and in the ferromagnetic regime ($\Delta=-1,-2$).
While we show in Fig.\ \ref{fig:2}(a) that we reproduce the results
by Alvarez and Gros\cite{gros02} for system sizes of $N\leq 14$, our
analysis extends this case to $N\leq 18$. This is due to exploiting both
conservation of total $S^z$ and momentum $k$ in the exact diagonalization.
By comparing with the curve obtained from the Bethe ansatz\cite{kluemper02}
(solid line in Fig.\ \ref{fig:2}(a)) it can be seen that for a system of
$N=18$ sites the thermodynamic limit is reached for temperatures around $T\gtrsim 0.3 J $ for $\Delta=1$. At low
temperatures $D_{\mathrm{th}}(T)$ is exponentially suppressed due
to the finite-size gap in the case of an even number of sites
and divergent for an odd number. The latter is due to the degeneracy
of the ground state in case of odd-numbered systems. \\
\indent For the  gapped, antiferromagnetic case we choose  $\Delta=10$, shown in Fig.\ \ref{fig:2} (b),
where the finite-size effects of the   two-spinon gap are small.
The data for $D_{\mathrm{th}}$ are convergent for
$T\gtrsim 3J$, but
substantial finite-size effects are still present in the vicinity
of the maximum, i.e., at small temperatures compared  to the two-spinon gap $8.055126 J$\cite{cloizeaux66}.
At low temperatures the thermal Drude weight $D_{\mathrm{th}}(T)$
is expected  to be exponentially suppressed  in the thermodynamic limit.
In the inset of Fig.\ \ref{fig:2} (b)   $D_{\mathrm{th}}(T)$ is plotted logarithmically versus $1/T$.
 If one fits $D_{\mathrm{th}}(T)\sim \mbox{exp}(-\delta/T)$
 to the numerical data at low temperatures\cite{T_range}  one finds $\delta=8.056 J$ for
$N=18$ sites with similar values of $\delta$ found for other $N$.
This compares well to the two-spinon gap\cite{cloizeaux66}.
Hence we conclude that mainly the elementary excitations
contribute to the thermal conductivity at low temperatures.
\\
\indent
In  the ferromagnetic regime ($\Delta \leq -1$) (results are shown for $\Delta=-1$
and $\Delta=-2$ in Fig.\ \ref{fig:2}(c))
our main observation is that convergence with $N$ is very good at all temperatures.
For example, we find for $\Delta=-2$ that   the relative difference
between the finite-size data for
$N=16$ sites and $N= 17$ is negligibly small, namely
$|D_{\mathrm{th}}^{N=17}(T)-D_{\mathrm{th}}^{N=16}(T)|/D_{\mathrm{th}}^{N=17}(T)<0.008$ for $T> 0.05J$.
If one extracts  $\delta$ from a fit of $D_{\mathrm{th}}(T)\sim \mathrm{exp}(-\delta/T)$ to the
numerical data\cite{T_range} for $\Delta=-2$, we find   $\delta\approx 0.97 J$ which
 coincides with the one-triplet gap $-(\Delta+1)J$ that
can easily be obtained from a spin-wave computation.
The fast convergence  is  even more remarkable for the case $\Delta=-1$
where the numerical data are consistent with $D_{\mathrm{th}}(T)\sim T$ at low temperatures.
In addition,  there is no qualitative difference between even- and odd-numbered systems due to the
ferromagnetic nature of the interaction for $\Delta\leq-1$. \\
%---------------------------
%---------------------------
\indent {\bf Frustrated chain - } Now we turn to the case of nonzero frustration.
Since $\lbrack H,j \rbrack\not= 0$ here,  care has to be  taken about off-diagonal matrix-elements of
$j$ if degeneracies occur.
However, since we use classification by momentum $k$ and $S^z$ degeneracies are lifted and
do not play a crucial role.
In Fig.\ \ref{fig:3} we
show $D_{\mathrm{th}}(T)$ obtained numerically from Eqs.\ (\ref{eq:10}) and (\ref{eq:13})
 for even system sizes with $N\leq 18$ and
$\alpha=0.35, \Delta=1$. A central result of this paper is that, while we
observe a finite Drude weight at temperatures $T>0$ and all system sizes investigated,
we still find a substantial reduction of the Drude weight with increasing system size at high $T$.
This is in sharp contrast to the $XXZ$ model,
where finite-size effects are small at high temperatures
(see e.g.\ Fig.\ \ref{fig:2} (a)).
These observations clearly point to a vanishing of the Drude weight in the thermodynamic limit for $\alpha=0.35$.
However,  the question of dissipationless thermal transport
at arbitrary $\alpha>0$ remains to be studied in more detail. \\
  \begin{figure}[t]
\centerline{\epsfig{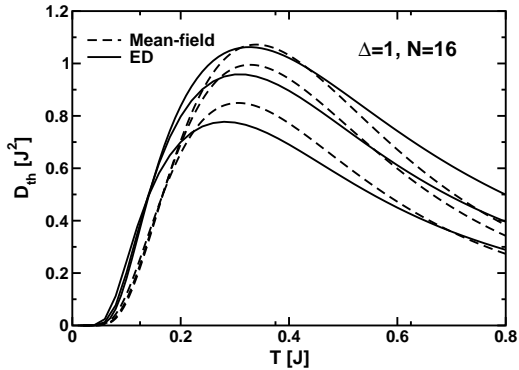}}
\caption{Comparison between exact diagonalization (solid lines) and mean-field approximation (dashed lines) on finite systems with $N=16$ sites
for various values of $\alpha=0,0.05,0.15$ (top to bottom) and $\Delta=1$.\label{fig:4}}
\end{figure}
\indent
Finally, we compare our mean-field approach with numerical results on finite
systems and with nonzero frustration.
In Fig.\ \ref{fig:4} we present the thermal Drude weight of systems with $N=16$
sites for $\alpha=0,0.05,0.15$ and $\Delta=1$, i.e., in the gapless regime.
As is obvious from this figure, there is a good
agreement between ED and the mean-field approach regarding the
temperature dependence of the thermal Drude weight.
The general features of $D_\mathrm{th}(T)$
are a reduction of the absolute value of $D_{\mathrm{th}}$ on
increasing $\alpha$, a shift of the position of the maximum to lower
temperatures and thus a crossing of the curves for different $\alpha$ at low
temperatures which are present in both the ED and mean-field results.
Deviations at high temperatures are again understandable due to the neglect of
many-particle excitations in the effective one-particle picture. \\
%----------------------------------------------
\indent {\bf Conclusion - }  We performed a detailed analysis of the thermal
Drude weight for anisotropic and frustrated spin-$1/2$-Heisenberg chains by
using mean-field theory, bosonization and ED. In the case of the $XXZ$ model
we demonstrated the applicability of these techniques for computing the
temperature dependence of the thermal Drude weight. Using ED we obtained
results on finite systems of $N\leq 18$ sites for arbitrary values of the
anisotropy $\Delta$. In the ferromagnetic regime ($\Delta\leq -1$) of the
$XXZ$ chain the numerical data converge to the thermodynamic limit at
arbitrary temperature for moderately small system sizes ($N \approx 18$).
The analytical results compare well with the Bethe ansatz\cite{kluemper02}
in the gapless regime of the $XXZ$ model and to our numerics in the case of
the frustrated chain on finite systems.
Our numerical data at $\alpha=0.35$
mark a clear difference between the integrable $XXZ$ case and the
nonintegrable one at $\alpha=0.35$: while in the former case the
Drude weight remains finite in the thermodynamic limit we have clear
indications for a vanishing Drude weight at high temperatures in the latter case.
Extended analysis of these findings for frustrated and dimerized chains
and spin ladders will be the subject of a forthcoming paper.\\
 %----------------------------------------------
\indent {\bf Acknowledgments - }
This work was supported by the DFG, Schwer\-punkt\-programm 1073, and by
a DAAD-ANTORCHAS exchange program. We acknowledge helpful discussions with
J.V.\ Alvarez, B.\ B\"uchner, C.\ Gros, and C.\ Hess.


%--------------------------------------------------------

\begin{thebibliography}{99}
\bibitem{current} In the absence of a chemical potential the energy-current
    operator is identical to the heat-current operator.
\bibitem{T_range} $D_{\mathrm{th}}\sim\mathrm{exp}(-\delta/T)$ was fitted to
    $D_{\mathrm{th}}(T)$ for $0.35 J\leq T\leq 0.85J$ ($\Delta=10$) and
    $T\leq 0.2J$ ($\Delta=-2$).
\end{thebibliography}
\end{document}